\documentclass[twoside]{article}

\usepackage{amsmath,amsthm,amssymb}

\allowdisplaybreaks[3]

\newtheorem{theorem}{Theorem}
\newtheorem{lemma}{Lemma}

\newtheorem{proposition}{Proposition}
\newtheorem{corollary}{Corollary}

\begin{document}

\title{On convergence of dynamics of hopping particles to a birth-and-death process in continuum}

\author{Dmitri Finkelshtein\thanks{Institute of Mathematics, National Academy of Sciences
of Ukraine, 3 Tereshchenkivska Str., Kiev 01601, Ukraine ({\tt
fdl@imath.kiev.ua})}, \and Yuri Kondratiev\thanks{ Fakult\"at f\"ur
Mathematik, Universit\"at Bielefeld, Postfach 10 01 31, D-33501
Bielefeld, Germany; Department of Mathematics, University of
Reading, U.K.; BiBoS, Univ.\ Bielefeld, Germany ({\tt
kondrat@mathematik.uni-bielefeld.de})}, \and Eugene
Lytvynov\thanks{Department of Mathematics, Swansea University,
Singleton Park, Swansea SA2 8PP, U.K. ({\tt
e.lytvynov@swansea.ac.uk})}}

\makeatletter

\renewcommand{\@oddhead}{\thepage \hfil \small D. Finkeleshtein, Y. Kondratiev and E. Lytvynov \hfil}

\renewcommand{\@evenhead}{\hfil \small Convergence of dynamics in continuum \hfil \thepage}

\makeatother

\maketitle

\begin{abstract}
We show that some classes of birth-and-death processes in continuum
(Glauber dynamics) may be derived as a scaling limit of a dynamics
of interacting hopping particles (Kawasaki dynamics)
\end{abstract}

{\small {\bf Keywords:} Continuous system, Gibbs measure,
birth-and-death process in continuum (Glauber dynamics), dynamics of
hopping particles (Kawasaki dynamics), scaling limit}

{\small {\bf Mathematical Subject Classification (2000):} 60K35,
60J75, 60J80, 82C21, 82C22}

\section{Preliminaries}

This letter deals with two classes of stochastic dynamics of
infinite particle systems in continuum. Let $\Gamma$ denote the
space of all locally finite subsets of $\mathbb R^d$, $d\in\mathbb
N$. This space is called the configuration space. Elements of
$\Gamma$  are called configurations, and each point of a
configuration represents position of a particle. We endow $\Gamma$
with the vague topology, i.e., the weakest topology in $\Gamma$ with
respect to which every mapping of the form
$\Gamma\ni\gamma\mapsto\langle
f,\gamma\rangle:=\sum_{x\in\gamma}f(x)$, with $f\in C_0(\mathbb
R^d)$, is continuous. Here $C_0(\mathbb R^d)$ is the space of all
real-valued functions on $\mathbb R^d$ with compact support. We
denote by $\mathcal B(\Gamma)$ the Borel $\sigma$-algebra in
$\Gamma$.

A dynamics of hopping particles (Kawasaki dynamics) is a Markov process on $\Gamma$ whose generator is given (on an appropriate set of functions on $\Gamma$) by
$$ (L_{\mathrm K}F)(\gamma)=\sum_{x\in\gamma}\int_{\mathbb R^d}dy\, c(x,y,\gamma\setminus x)(F(\gamma\setminus x\cup y)-F(\gamma)).$$
Here and below, for simplicity of notations, we just write $x,y$ instead of $\{x\},\{y\}$. The function $c(x,y,\gamma\setminus x)$ describes the rate at which a particle  $x$ of configuration $\gamma$ jumps to $y$, taking into account the rest of configuration, $\gamma\setminus x$.

A birth-and-death process in continuum (Glauber dynamics) is a Markov process on $\Gamma$ with generator
\begin{multline*} (L_{\mathrm G}F) (\gamma)=\sum_{x\in\gamma}d(x,\gamma\setminus x)(F(\gamma\setminus x)-F(\gamma))\\ \text{}+\int_{\mathbb R^d} dy\, b(y,\gamma)(F(\gamma\cup y)-F(\gamma)).\end{multline*} Here $d(x,\gamma\setminus x)$ describes the rate at which a particle $x$ of configuration $\gamma$ dies, whereas $b(x,\gamma)$ describes the rate at which, given configuration $\gamma$, a new particle is born at $y$. Fore some constructions and discussions of Glauber and Kawsaki dynamics in continuum, see \cite{BCC,HS,GK,KKZ,KL,KLR,KLR2,P} and the references therein.

The aim of this letter it to show that, in many cases, a birth-and-death process may be interpreted as a limiting dynamics of hopping particles. We will restrict out attention  to the case where the rate $c$ of the Kawasaki dynamics is given by
$$c(x,y,\gamma\setminus x)=a(x-y)\exp[E^{\phi^-}(x,\gamma\setminus x)-E^{\phi^+}(y,\gamma\setminus x)].$$
Here $a$ and $\phi^{\pm}$ are even functions on $\mathbb R^d$ (e.g.\ $a(-x)=a(x)$), $a$ is bounded, $a\ge0$,  $\int_{\mathbb R^d}a(x)\,dx=1$, and for $x\in\mathbb R^d$ and $\gamma\in\Gamma$, $$E^{\phi^{\pm}}(x,\gamma):=\sum_{y\in\gamma}\phi^{\pm}(x-y),$$ provided the sum converges absolutely.  Thus, $c(x,y,\gamma\setminus x)$ is a product of three terms: the term $e^{E^{\phi^-}(x,\gamma\setminus x)}$ describes the rate at which a particle $x\in\gamma$ jumps, the term $e^{-E^{\phi^+}(y,\gamma\setminus x)}$ describes the rate at which this particle lands at $y$, and finally the term $a(x-y)$ gives the distribution of an individual jump.

We now produce the following scaling of this dynamics. For each $\varepsilon>0$, we define $a_{\varepsilon}(x):=\varepsilon ^d a(\varepsilon x)$.
We clearly have that $\int_{\mathbb R^d}a_\varepsilon(x)\,dx=1$.
Let $c_\varepsilon$ denote the $c$ coefficient in which function $a$ is replaced by $a_\varepsilon$, and let $L_\varepsilon$ denote the corresponding $L_{\mathrm K}$ generator. Letting $\varepsilon\to0$, we may suggest that only jumps of infinite length will survive, i.e., jumps from a point to `infinity', and jumps from `infinity' to a point. Thus, we expect to arrive at a birth-and-death process. To make our suggestion more explicit, we proceed as follows.

\section{Convergence of the generator of the scaled evolution of correlation functions}

For simplicity, we assume, in this section,  that the functions $\phi^{\pm}$ are from $C_0(\mathbb R^d)$. Then $E^{\phi^{\pm}}(x,\gamma)$ are well defined for each $x\in\mathbb R^d$ and $\gamma\in\Gamma$.

Let us briefly recall some basic facts of harmonic analysis on the configuration space, see \cite{FKO,KK} for further detail. Let $\Gamma_0$ denote the space of all finite configurations  in $\mathbb R^d$, i.e., $\Gamma_0=\bigcup_{n=0}^\infty\Gamma^{(n)}$, where $\Gamma^{(n)}$ is the space of all $n$-point configurations in $\mathbb R^d$. Clearly, $\Gamma_0\subset\Gamma$, and we define
$\mathcal B(\Gamma_0)$ and $\mathcal B(\Gamma^{(n)})$ as the trace $\sigma$-algebra of $\Gamma$ on $\Gamma_0$ and $\Gamma^{(n)}$, respectively. For a function $G:\Gamma_0\to\mathbb R$, we define a function
$(KG)(\gamma):=\sum_{\eta\Subset\gamma}G(\eta)$, $ \gamma\in\Gamma$, provided the summation makes sense. Here $\eta\Subset\gamma$ means that    $\eta$ is a finite subset of $\gamma$.

Let $\mu$ be a  probability measure
 on $(\Gamma,\mathcal B(\Gamma))$.
Then there exists a unique measure $\rho$ on $(\Gamma_0,\mathcal B(\Gamma_0))$ satisfying
$$ \int_{\Gamma}(KG)(\gamma)\,\mu(d\gamma)=\int_{\Gamma_0}G(\eta)\,\rho(d\eta)$$ for each measurable function $G:\Gamma_0\to[0,\infty)$.
The measure $\rho$ is called the correlation measure of $\mu$. Further, denote by $\lambda$ the Lebesgue--Poisson measure on $\Gamma_0$, i.e.,
$$\lambda=\delta_{\varnothing}+\sum_{n=1}^\infty\frac1{n!}\,dx_1\cdots dx_n.$$
Here $\delta_{\varnothing}$ is the Dirac measure
with mass at $\varnothing$, and $dx_1\cdots dx_n$
is the Lebesgue measure on $\Gamma^{(n)}$, which is naturally defined  on this space. Assume that the correlation measure $\rho$ of $\mu$ is absolutely continuous with respect to $\lambda$. Then $k:=\frac{d\rho}{d\lambda}$ is called the correlation functional of $\mu$.
For a given correlation functional $k$, the corresponding Ursell functional $u:\Gamma_0\to\mathbb R$ is defined through the formula 
$k(\eta)=\sum_{\pi\in\mathcal P(\eta)}u_{\pi}(\eta)$,
where $\mathcal P(\eta)$ denotes the set of all partitions of $\eta$, and given a partition $\pi=\{\eta_1,\dots,\eta_k\}$ of $\eta$, $ u_\pi(\eta):=u(\eta_1)\dotsm u(\eta_k)$.
Recall also that a function $G:\Gamma_0\to\mathbb R$ is called translation invariant if, for each $x\in\mathbb R^d$, $G(\eta_x)=G(\eta)$ for all $\eta\in \Gamma_0$, where $\eta_x$ denotes the configuration $\eta$ shifted by vector $x$, i.e., $\eta_x:=\{y+x\mid y\in\eta\}$. Clearly, the correlation functional $k$ is translation invariant if and only if the corresponding Ursell functional $u$ is translation invariant.

If $k$ is the correlational functional of a probability measure $\mu$ on $\Gamma$, we denote
$$k^{(n)}(x_1,\dots,x_n):=k(\{x_1,\dots,x_n\}),\quad n\in\mathbb N,$$ and analogously we define $u^{(n)}$. The $(k^{(n)})_{n=1}^\infty$ and $(u^{(n)})_{n=1}^\infty$ are called the correlation and Ursell functions of $\mu$, respectively. Note that,  if $k$ is translation invariant, then $k^{(1)}=u^{(1)}$ is a constant.

For a function $f:\mathbb R^d\to\mathbb R$, we define $e_\lambda(f,\eta):=\prod_{x\in\eta}f(x)$, $\eta\in\Gamma_0$, where $\prod_{x\in\varnothing}f(x):=1$. Further, let $\varphi:\mathbb R^d\to\mathbb R$. Then
$$ (Ke_\lambda(e^\varphi-1,\cdot))(\gamma)=e^{\langle\varphi,\gamma\rangle},$$
so that
\begin{equation}\label{daftiuyf}\int_\Gamma e^{\langle\varphi,\gamma\rangle}\,\mu(d\gamma)=
\int_{\Gamma_0}e_\lambda(e^\varphi-1,\eta)k(\eta)\,\lambda(d\eta),\end{equation}
under some proper conditions on $\varphi$ and $k$, see e.g. \cite{KK}.

Assume that $L$ is a Markov generator on $\Gamma$. Denote $\hat L:=K^{-1}LK$, i.e., $\hat L$ is the operator acting on functions on $\Gamma_0$ which satisfies $K\hat LG=LKG$. Denote by $\hat L^*$ the dual operator of $\hat L$ with respect to the Lebesgue--Poisson measure $\lambda$:
$$\int_{\Gamma_0} (\hat LG)(\eta)k(\eta)\,\lambda(d\eta)=\int_{\Gamma_0}G(\eta)(\hat L^*k)(\eta)\,\lambda(d\eta).$$  Assume now that a Markov process on $\Gamma$ with generator $L$ has initial distribution $\mu_0$. Denote by $\mu_t$ the distribution of this process at time $t>0$. Assume that, for each $t\ge0$, $\mu_t$ has correlation functional $k_t$. Then, at least at an informal level, one sees that the evolution of $k_t$ is described by the equation
$ \partial k_t/\partial t=\hat L^*k_t$,
so that $\hat L^*$ is the generator of evolution of correlation functionals.

In the case where $L=L_\varepsilon$, we proceed as follows, First we write $L_\varepsilon=L_\varepsilon^-+ L_\varepsilon^+$, where
\begin{align*}
&(L_\varepsilon^-F)(\gamma)=\sum_{x\in\gamma}\int_{\mathbb R^d}dy\,a_\varepsilon(x-y)r(x,y,\gamma\setminus x)(F(\gamma\setminus x)-F(\gamma)),\\
&(L_\varepsilon^+)(\gamma)=\sum_{x\in\gamma}\int_{\mathbb R^d}dy\,a_\varepsilon(x-y)r(x,y,\gamma\setminus x)(F(\gamma\setminus x\cup y)-F(\gamma\setminus x)).
\end{align*}
Here, $r(x,y,\gamma\setminus x):=\exp[E^{\phi^-}(x,\gamma\setminus x)-E^{\phi^+}(y,\gamma\setminus x)]$. We also set
\begin{align*}
&(L_0^-F)(\gamma)=\sum_{x\in\gamma}
\exp[E^{\phi^-}(x,\gamma\setminus x)](F(\gamma\setminus x)-F(\gamma)),\\
&(L_0^+F)(\gamma)=\int_{\mathbb R^d}dy\,
\exp[-E^{\phi^+}(y,\gamma)](F(\gamma\cup y)-F(\gamma)).
\end{align*}

\begin{theorem}\label{cfttxd}
Let $k$ be the correlation functional of a probability measure
$\mu$ on $(\Gamma,\mathcal B(\Gamma))$, and let $u$ be the
corresponding Ursell functional. Assume that the following
conditions are satisfied:

\begin{enumerate}

\item[{\rm i)}] $k$ fulfills the  bound $k(\eta)\le (|\eta|!)^{s}C^{|\eta|}$, $\eta\in\Gamma_0$, for some $0\le s<1$ and $C>0$. Here $|\eta|$
denotes the cardinality of set $\eta$.

\item[{\rm ii)}] $k$  is translation invariant.

\item[{\rm iii)}] The measure $\mu$ has a decay of correlations in the sense that, for any $n,m\in\mathbb N$, $a\in\mathbb R^d$, $a\ne0$, and
$\{x_1,\dots,x_{n+m}\}\in\Gamma^{(n+m)}$,
$$ u(\{x_1,\dots,x_n,x_{n+1}+(a/\varepsilon),\dots,
x_{n+m}+(a/\varepsilon)\})\to 0\quad\text{as }\varepsilon\to0.$$
\end{enumerate}
Then, for each $\eta\in\Gamma_0$,
$$
(\hat L_\varepsilon^-{}^*k)(\eta)\to
c^-(k)(\hat L_0^-{}^*k)(\eta),
\quad
(\hat L_\varepsilon^+{}^*k)(\eta)\to
c^+(k)(\hat L_0^+{}^*k)(\eta),
$$
where
\begin{align}
c^-(k):=&\int_{\Gamma_0}\lambda(d\xi)\,e_\lambda(e^{-\phi^+}-1,\xi)k(\xi),\notag\\
c^+(k):=&\int_{\Gamma_0}\lambda(d\xi)\,e_\lambda(e^{\phi^-}-1,\xi)k(\xi\cup0).\label{cvii}
\end{align}
\end{theorem}

{\it Proof}. A straightforward calculation (see \cite{FKO}) shows
that
\begin{align}
(\hat L_\varepsilon^-{}^*k)(\eta)&=-\sum_{x\in\eta}\int_{\mathbb R^d}dy\, a_\varepsilon(x-y)r(x,y,\eta\setminus x)\notag\\
&\quad\times\int_{\Gamma_0 }\lambda(d\xi)\,k(\xi\cup\eta)e_\lambda(e^{\phi^-(x-\cdot)-\phi^+(y-\cdot)}-1,\xi),\label{esresa}\\
(\hat L_\varepsilon^+{}^*k)(\eta)&=\sum_{y\in\eta}
\int_{\mathbb R^d}dx\, a_\varepsilon(x-y)r(x,y,\eta\setminus y)\notag\\
&\quad\times\int_{\Gamma_0 }\lambda(d\xi)\, k(\xi\cup(\eta\setminus y)\cup x)e_\lambda(e^{\phi^-(x-\cdot)-\phi^+(y-\cdot)}-1,\xi),\notag\\
(\hat L_0^-{}^*k)(\eta)&=-\sum_{x\in\eta}\exp[E^{\phi^-}(x,\eta\setminus x)]\notag \\
&\quad\times\int_{\Gamma_0}\lambda(d\xi)\,e_\lambda(e^{\phi^-(x-\cdot)}-1,\xi)k(\eta\cup\xi),\notag\\
(\hat L_0^+{}^*k)(\eta)&=\sum_{y\in\eta}\exp[-E^{\phi^-}(y,\eta\setminus y)]\notag\\
&\quad\times
\int_{\Gamma_0}\lambda(d\xi)\,e_\lambda(e^{-\phi^+(y-\cdot)}-1,\xi)k((\eta\setminus y)\cup\xi).\notag
\end{align}

We will now briefly explain the convergence of
$(\hat L_\varepsilon^-{}^*k)(\eta)$ (the case of $(\hat L_\varepsilon^+{}^*k)(\eta)$ can be dealt with analogously).
From  (3) 
 and the definition of $\lambda$,
by making a change of variable, we easily have:
\begin{align*}
&(\hat L_\varepsilon^-{}^*k)(\eta)=-\sum_{x\in\eta}\int_{\mathbb R^d}dy\,a(y)r(x,(y/\varepsilon)+x,\eta\setminus x)\sum_{n=0}^\infty\frac1{n!}\sum_{k=0}^n
\binom nk\\
&\times\int_{(\mathbb R^d)^n}du_1\dotsm du_n
\prod_{i=1}^k\big(e^{-\phi^+((y/\varepsilon)+x-u_i)}(e^{\phi^-(x-u_i)}-1)\big)\\
&\times \prod_{j=k+1}^{n}(e^{-\phi^+((y/\varepsilon)+x-u_j)}-1)k(\xi\cup\{u_1,\dots,u_n\})\\
&=-\sum_{x\in\eta}\int_{\mathbb R^d}dy\,a(y)r(x,(y/\varepsilon)+x,\eta\setminus x)\sum_{n=0}^\infty\sum_{k=0}^n \frac1{k!(n-k)!}\\
&\times\int_{(\mathbb R^d)^n}du_1\dotsm du_n
\prod_{i=1}^k\big(e^{-\phi^+((y/\varepsilon)-u_i)}(e^{\phi^-(u_i)}-1)\big)\prod_{j=k+1}^{n}(e^{-\phi^+(u_j)}-1)\\
&\times k(\xi\cup\{u_1+x,\dots,u_k+x,u_{k+1}+x+(y/\varepsilon),\dots,u_n+x+(y/\varepsilon)\}).
\end{align*}
Next, represent the correlation functionals in the above expression through a sum of Ursell functionals. Using the dominated convergence theorem  and conditions i) and iii), we see that, in the limit, all the Ursell functionals containing at least
one point from $\xi\cup\{u_1+x,\dots,u_k+x\}$ and at least one point from $\{u_{k+1}+x+(y/\varepsilon),\dots,u_n+x+(y/\varepsilon)\}$ will vanish, and by virtue of ii), we conclude that $(\hat L_\varepsilon^-{}^*k)(\eta)$ converges to
\begin{align*}
&-\sum_{x\in\eta}\int_{\mathbb R^d}dy\,a(y)
\exp[E^{\phi^-}(x,\eta\setminus x)]
\sum_{n=0}^\infty\sum_{k=0}^n \frac1{k!(n-k)!}\notag\\
&\times\int_{(\mathbb R^d)^n}du_1\dotsm du_n
\prod_{i=1}^k(e^{\phi^-(x-u_i)}-1)\prod_{j=k+1}^{n}(e^{-\phi^+(u_j)}-1)\notag\\
&\times k(\xi\cup\{u_1,\dots,u_k\})k(\{u_{k+1},\dots,u_n\}),
\end{align*}
from where the statement follows.\quad $\square$

From Theorem~\ref{cfttxd}, we can make the following conclusion. Assume that a  dynamics of hopping particle with Markov generator $L_{\mathrm K}$ has initial distribution $\mu_0$. Let  $\mu_t$ be the distribution of this process at time $t>0$.
Assume that, for each $t\ge0$, $\mu_t$ has correlation functional $k_t$ which satisfies conditions i)--iii) of Theorem~\ref{cfttxd}. Further assume that $c^{\pm}(k_t)$, $t\ge0$, given through (2) 
remain constant. Then, we can expect  that the scaled dynamics of hopping particles converges to a birth-and-death process with generator $L_0:=c^-(k_0)L_0^-+c^+(k_0)L_0^+$
and initial distribution $\mu_0$. We will discuss below two cases where this statement can be proven rigorously (at least in the sense of convergence of the generators).

\section{Convergence of non-equilibrium free dyna\-mics} This case has been discussed in \cite{KLR2}, so here we will explain its connection with Theorem~\ref{cfttxd}.

Let $\Theta\in\mathcal B(\Gamma)$ be the set of those configurations $\gamma\in\Gamma$ for which there exist $\alpha\ge d$ and $K>0$ such that
\begin{equation} |\gamma\cap B(n)|\le Kn^\alpha,\quad\text{for all }n\in\mathbb N,\label{dtt}
\end{equation}
where $B(n)$ denotes the ball in $\mathbb R^d$
centered at 0 and of radius $n$. Note that the estimate (4) 
controls the growth of the number of particles of $\gamma$ at infinity.

Let $a\in S(\mathbb  R^d)$ (the Schwartz space of rapidly decreasing, infinitely differentiable functions on $\mathbb  R^d$). Consider
a random walk in $\mathbb R^d$ with transition kernel $Q(x,dy):=a(x-y)\,dy$. This is a Markov process in $\mathbb R^d$ with  generator
$$ (L^{(1)}f)(x)=\int_{\mathbb R^d}(f(y)-f(x))a(x-y)\,dy.$$ The corresponding Markov semigroup on $L^2(\mathbb R^d,dx)$ is then given by
\begin{equation}\label{12333} (p_tf)(x)=e^{-t}f(x)+\int_{\mathbb R^d}G(x-y)f(y)\,dy,\end{equation}
where $G$ is the inverse Fourier transform of
$e^{-t}(\exp[t(2\pi)^{d/2}\hat a]-1)$, where $\hat a$ is the Fourier transform of $a$.
(Note that we have normalized the direct and inverse Fourier transforms  so that they are unitary operators in $L^2(\mathbb R^d\to\mathbb C,dx)$.)
For any $\gamma\in\Theta$, consider a dynamics of independent particles which starts at $\gamma$ and such that each separate particle moves according to the semigroup $p_t$ (i.e., independent random walks in $\mathbb R^d$).
Then, this process has {\it
c\'adl\'ag} paths on $\Gamma$ and  a.s.\ it never leaves $\Theta$, cf.\ \cite{KLR2}.
The generator of the obtained Markov process on $\Theta$ is then given by
\begin{equation}\label{fttfytyfdw} (L_{\mathrm K}F)(\gamma)=\sum_{x\in\gamma}\int_{\mathbb R^d}dy\, a(x-y)(F(\gamma\setminus x\cup y)-F(\gamma)), \end{equation}
so that now $\phi^{\pm}=0$.

\begin{proposition}\label{hajy}
Let $\mu_0$ be a probability measure on $\Gamma$ whose correlation functional $k_0$ satisfies conditions i)--iii) of Theorem~\ref{cfttxd}, and $\mu_0(\Theta)=1$.  Consider the Markov process on $\Theta$ with the generator $L_{\mathrm K}$ given by (6) 
and with the initial distribution $\mu_0$.
Denote by $\mu_t$ the distribution of this process at time $t>0$.
Then, for each $t>0$,  $\mu_t$ has correlation functional $k_t$ which satisfies
conditions i)--iii) of Theorem~\ref{cfttxd}, and furthermore $c^-(k_t)=1$ and $c^+(k_t)=k_0^{(1)}$,
$t\ge0$.
\end{proposition}

{\it Proof.}  For each $f\in C_0(\mathbb R^d)$ and $t>0$, we have, by (1) and the construction of the process:
\begin{align}
&\int_\Theta \mu_t(d\gamma)e^{\langle f,\gamma\rangle}=\int_\Theta \mu_0(d\gamma)\prod_{x\in\gamma}(p_te^f)(x)\notag\\
&=\int_{\Gamma_0}\lambda(d\eta)k_0(\eta)\prod_{x\in\eta}(p_t(e^f-1))(x)\notag\\
&=1+\sum_{n=1}^\infty\frac1{n!}\int_{(\mathbb R^d)^n}
dx_1\dotsm dx_n\, k^{(n)}(x_1,\dots,x_n)\prod_{i=1}^n
(p_t(e^f-1))(x_i)\notag\\
&=1+\sum_{n=1}^\infty\frac1{n!}\int_{(\mathbb R^d)^n}
dx_1\dotsm dx_n\, (p_t^{\otimes n}k^{(n)})(x_1,\dots,x_n)\prod_{i=1}^n
(e^{f(x_i)}-1).\notag
\end{align}
Therefore, $\mu_t$ has correlation functional $k_t$, and furthermore $k_t^{(n)}=p_t^{\otimes n}k_0^{(n)}$. The latter equality, in turn, implies that
$u_t^{(n)}=p_t^{\otimes n}u_0^{(n)}$.
From here  it easily follows that, for each $t>0$, $\mu_t$  satisfies assumptions i)--iii) of Theorem~\ref{cfttxd}. Furthermore, by  (2),
\begin{align*}
c^-(k_t)&=k_t(\varnothing)= 1,\\
c^+(k_t)&=k_t(\{0\})=k_t^{(1)}=p_tk_0^{(1)}=k_0^{(1)}.\quad \square
\end{align*}

Thus, according to Section~2, we expect that the scaled free dynamics with initial distribution $\mu_0$ converges to the birth-and-death process with  generator
\begin{equation}\label{agcsy} (L_0F)(\gamma)=\sum_{x\in\gamma}(F(\gamma\setminus x)-F(\gamma))+k_0^{(1)}\int_{\mathbb R^d}dy\,(F(\gamma\cup y)-F(\gamma))\end{equation} and initial distribution $\mu_0$.
This dynamics can be constructed as follows, cf.\  \cite{KLR2,Surgailis2}.
For each $\gamma\in\Theta$, denote by $P_{\gamma}$ the law of a process on $\Theta$ which is at $\gamma$ at time zero, and after  this,  points of $\gamma$ randomly die, independently of each other, so that the probability that at time $t>0$ a particle $x\in\gamma$ is still alive is equal to
 $e^{- t}$.
 Next, let $\pi$ denote the Poisson point process in  $\mathbb R^d\times(0,\infty)$ with the intensity measure
$k_0^{(1)}\, dx\,dt$. The measure $\pi$ is concentrated on configurations $\widehat\gamma=\{(x_n,t_n)\}_{n=1}^\infty$ in $\mathbb R^d\times(0,\infty)$ such that $\{x_n\}_{n=1}^\infty\in\Theta$,
$0<t_1<t_2<\cdots$, and $t_n\to\infty$ as $n\to\infty$. For any such configuration, we denote by $P{}_{\widehat\gamma}$ the law of a process on $\Theta$ such that at time $t=0$, the configuration is empty, and then at each time $t_n$, $n\in\mathbb N$, a new particle is born at $x_n$, and after time $t_n$ this particle randomly dies, independently of the other particles, so that at time $s>t_n$ the probability that the particle is still alive is $e^{- (s-t_n)}$. Finally, the law of the process with generator
(7) 
and initial distribution $\mu_0$ is given by
$$ \int \mu_0(d\gamma)P_\gamma*\int\pi(d\widehat\gamma)P{}_{\widehat\gamma}.$$
Here $*$ stays for convolution of measures, see \cite{KLR2} for
details.

We will use $\ddot\Gamma$ to denote the space of multiple
configurations over $\mathbb R^d$ equipped with the vague topology,
see e.g.\ \cite{Kal} for details. Note that
$\Gamma\subset\ddot\Gamma$, and the trace $\sigma$-algebra of
$\mathcal B(\ddot\Gamma)$ on $\Gamma$ is $\mathcal B(\Gamma)$.

\begin{theorem}[\cite{KLR2}]\label{r66e}
Consider the stochastic process from Proposition~1 as taking values in $\ddot\Gamma$. Then, after scaling, this process converges, in the sense of weak convergence of finite-dimensional distributions, to the Markov process with the generator $L_0$ given by (7) 
 and with the initial distribution $\mu_0$.
\end{theorem}

Note that the limiting process also lives in $\Theta$, and we used the $\ddot\Gamma$ space only to identify the type of convergence.

For reader's convenience, let us explain the idea of the proof of Theorem~2. Fix arbitrary $0=t_0<t_1<t_2<\dots<t_n$, $n\in\mathbb N$, and denote by
$\mu^\varepsilon_{t_0,t_1,\dots,t_n}$, $\varepsilon\ge0$, the corresponding finite-dimensional distribution of the initial process scaled by $\varepsilon>0$, and that of the limiting process if $\varepsilon=0$, respectively.   Then, by \cite{Kal}, the statement of the theorem is equivalent to staying that, for any non-positive $f_0,f_1,\dots,f_n\in C_0(\mathbb R^d)$,
\begin{multline}\label{ycdtr}
\int_{\Theta^n}\exp\bigg[ \sum_{i=0}^n \langle f_i,\gamma\rangle\bigg]d\mu^\varepsilon_{t_0,t_1,\dots,t_n} (\gamma_0,\gamma_1,\dots,\gamma_n) \\
\to \int_{\Theta^n}\exp\bigg[ \sum_{i=0}^n \langle f_i,\gamma\rangle\bigg]d\mu^0_{t_0,t_1,\dots,t_n} (\gamma_0,\gamma_1,\dots,\gamma_n)
\quad\text{as }\varepsilon\to0.
\end{multline}
For $\varepsilon>0$, denote by $p_t^\varepsilon(x,dy)$ the transition probability of the Markov semigroup (4) 
scaled by $\varepsilon$. Set
\begin{multline*} g^\varepsilon(x):=e^{f_0(x)}\int_{\mathbb R^d}p_{t_1}^\varepsilon(x,dx_1)\int_{\mathbb R^d}p^\varepsilon
_{t_2-t_1}(x_1,dx_2)\\
\times\dotsm\times  \int_{\mathbb R^d}p^\varepsilon
_{t_n-t_{n-1}}(x_{n-1},dx_n)\prod_{i=1}^n e^{f_i(x_i)},\quad x\in\mathbb R^d.\end{multline*}
Then, by (1) and the construction of the process, the first integral in (8) 
(with $\varepsilon>0$) is equal to
\begin{multline*} \int_{\Theta}\prod_{x\in\gamma}g^\varepsilon(x)\,\mu_0(d\gamma)\\
=1+\sum_{n=1}^\infty\frac1{n!}\int_{(\mathbb R^d)^n}\prod_{i=1}^n (g^\varepsilon(x_i)-1)
k_0^{(n)}(x_1,\dots,x_n)\,dx_1\dotsm dx_n.
\end{multline*}
In the above integrals, one represents the correlation functions through the Ursell functions, makes a change of variables under the sign of integral, and after a careful analysis of the obtained expression, one takes its limit as $\varepsilon\to0$. Finally, one shows that the obtained limit is indeed equal to the second integral in
(8)
.

\section{Convergence of equilibrium Kawasaki dyna\-mics of interacting particles}
In this section, we will consider equilibrium dynamics of interacting particles having a Gibbs measure as an equilibrium measure. Our result will  extend that of \cite{FKL}, where just one special case of such a dynamics was considered
(see also \cite{LP}). We start with a description of the class of Gibbs measures we are going to use.

A pair potential is a Borel-measurable function $\phi:\mathbb R^d \to\mathbb R\cup\{+\infty\}$
such that $\phi(-x)=\phi(x)\in\mathbb R$ for all $x\in\mathbb R^d\setminus\{0\}$.
For $\gamma\in\Gamma$ and $x\in\mathbb R^d\setminus\gamma$, we define a relative
energy of interaction between a particle at $x$ and the
configuration $\gamma$ as
$E(x,\gamma):=\sum_{y\in\gamma}\phi(x-y)$, provided that the latter sum converges absolutely, and otherwise it is set to be $=\infty$.
A (grand canonical) Gibbs measure corresponding to
the pair potential $\phi$ and  activity $z>0$ is a probability measure $\mu$ on $(\Gamma,\mathcal B(\Gamma))$ which satisfies
the Georgii--Nguyen--Zessin identity:
\begin{equation}\label{GNZ}
\int_\Gamma\mu(d\gamma)\sum_{x\in\gamma}F(\gamma,x)
=\int_\Gamma \mu(d\gamma)\int_{\mathbb R^d} z\,dx \exp[- E(x,\gamma)]
F(\gamma\cup x,x)
\end{equation}
for any measurable function $F:\Gamma\times\mathbb R^d\to[0,+\infty)$. A pair potential $\phi$ is said to be stable if there exists $B\ge0$ such that, for any $\eta\in\Gamma_0$,
\begin{equation}\label{cft}\sum_{\{x,y\}\subset\eta}\phi(x-y)\ge-B|\eta|.\end{equation}
In particular, we then have $\phi(x)\ge-2B$, $x\in\mathbb R^d$. Next, we say that the condition of low activity--high temperature regime is fulfilled if \begin{equation}\label{yagyaug} \int_{\mathbb  R^d}|e^{-\phi(x)}-1|z\,dx<(2e^{1+2B})^{-1},\end{equation}
where $B$ is as in (10)
. A classical result of Ruelle \cite{wewewe,Ru69} says that, under the assumption of stability and low activity--high temperature regime, there exists a Gibbs measure $\mu$ corresponding to $\phi$ and $z$,
and this measure has correlation functional which satisfies conditions i)--iii) of Theorem~\ref{cfttxd},
with $s=0$ in condition i) (which is then called the Ruelle bound).
Furthermore, the corresponding Ursell functions satisfy $u^{(n)}(0,\cdot,\dots,\cdot)\in L^1((\mathbb R^d)^{n-1},dx_1\dotsm dx_n)$  for each $n\ge2$.
In what follows, we will  assume that the potential $\phi$ is also bounded  from above outside some finite ball in $\mathbb R^d$
(which is always true for any realistic potential, since it should converge to zero at infinity).

We now fix arbitrary parameters $u,v\in[0,1]$, and assume that
\begin{equation}\label{ftdfyd}
\int_{\mathbb R^d}|\exp[(2(u\vee v)-1)\phi(x)]-1|\,dx<\infty.
\end{equation}
It can be easily shown that, if
$u,v\in[0,1/2]$, then  (12) 
 is a corollary of (11) 
 and the condition that
$\phi$ be  bounded   outside some finite ball.
Note that, even if $u\vee v\in(1/2,1]$,  condition (12) 
still admits potentials which have `weak' singularity at zero.

We  introduce the set $\mathcal FC_b(C_0(\mathbb R^d),\Gamma)$ of all functions of the form
$$\Gamma \ni \gamma \mapsto F(\gamma)=g(\langle f_1,\gamma \rangle,\dots,\langle f_N,\gamma \rangle),$$
where $N\in\mathbb N$, $f_1,\dots,f_N \in C_0(\mathbb R^d)$, and $g\in C_b(\mathbb R^N)$. Here  $C_b(\mathbb R^N)$
denotes the set of all continuous bounded functions on $\mathbb R^N$.
For each $F\in \mathcal FC_b(C_0(\mathbb R^d),\Gamma)$, we  define
\begin{multline}\label{ysdfrr} (L_{\mathrm K}F)(\gamma)=\frac12\sum_{x\in\gamma}\int_{\mathbb R^d}dy\, a(x-y)\big(
\exp[uE(x,\gamma\setminus x)-(1-v)E(y,\gamma\setminus x)]
\\\text{}+\exp[vE(x,\gamma\setminus x)-(1-u)E(y,\gamma\setminus x)] \big)(F(\gamma\setminus x\cup y)-F(\gamma)).\end{multline}
Note that the first addend in (13) 
corresponds to the choice of $\phi^-=u\phi$, $\phi^+=(1-v)\phi$, whereas the second addend corresponds to $\phi^-=v\phi$, $\phi^+=(1-u)\phi$. In the special case where $u=v$, we get
\begin{multline*} (L_{\mathrm K}F)(\gamma)=\sum_{x\in\gamma}\int_{\mathbb R^d}dy\, a(x-y)
\exp[uE(x,\gamma\setminus x)-(1-u)E(y,\gamma\setminus x)]\\ \times
(F(\gamma\setminus x\cup y)-F(\gamma)).\end{multline*}

By  \cite{KLR}, $(L_{\mathrm K},\mathcal FC_b(C_0(\mathbb R^d),\Gamma))$ is a Hermitian, non-negative operator in $L^2(\Gamma,\mu)$, and we denote by $(L_{\mathrm K},D(L_{\mathrm K}))$ its Friedrichs' extension. As shown in  \cite{KLR} by using the theory of Dirichlet forms, there exists a Markov process on $\Gamma$ with {\it c{\'a}dl{\'a}g\/} paths whose generator is   $(L_{\mathrm K},D(L_{\mathrm K}))$.
If we consider this process with initial distribution $\mu$, then it is an equilibrium process, i.e., it has distribution $\mu_t=\mu$ at any moment of time $t\ge0$. Thus, for each $t\ge0$, $\mu_t=\mu$ has correlation function which satisfies conditions i)--iii) of Theorem~\ref{cfttxd}.

\begin{lemma}Let $k$ denote the correlation function of the Gibbs measure $\mu$ under consideration.
Denote
$$ C_u:=\int_{\Gamma}\mu(d\gamma)\exp[-(1-u)\langle \phi,\gamma\rangle].$$
Then we have:
\begin{align}
&\int_{\Gamma_0}\lambda(d\xi)e_\lambda(e^{-(1-u)\phi}-1,\xi)k(\xi)=C_u,\label{hftgv}\\
&\int_{\Gamma_0}\lambda(d\xi)
e_\lambda(e^{u\phi}-1,\xi)k(\xi\cup0)
=zC_u.\label{vftyffty}\end{align}
\end{lemma}

{\it Proof.} Equality (14) 
follows from (1). Next, using (1), (9)
,  and translation invariance of $k$, we have, for each $f\in C_0(\mathbb R^d)$:
\begin{align*}
&\int_{\mathbb R^d}dx\,f(x)\int_\Gamma\mu(d\gamma)\exp[-(1-u)\langle\phi,\gamma\rangle]\\
&=\int_{\mathbb R^d}dx\,f(x)\int_\Gamma\mu(d\gamma)\exp[-(1-u)E(x,\gamma)]\\
&=z^{-1}\int_\Gamma\mu(d\gamma)\sum_{x\in\gamma}f(x)\exp[uE(x,\gamma\setminus x)]\\
&=z^{-1}\int_\Gamma\mu(d\gamma)\sum_{x\in\gamma}f(x)\sum_{\xi\Subset\gamma\setminus x}e_\lambda(e^{u\phi}-1,\xi)\\
&=z^{-1}\int_\Gamma\mu(d\gamma)
\sum_{\xi\Subset\gamma}\sum_{x\in\xi}
f(x)e_\lambda(e^{u\phi}-1,\xi\setminus x)\\
&=z^{-1}\int_{\Gamma_0}\lambda(d\xi)k(\xi)\sum_{x\in\xi}f(x)e_\lambda(e^{u\phi}-1,\xi\setminus x)\\
&=z^{-1}\int_{\Gamma_0}\lambda(d\xi)\int_{\mathbb R^d}dx\, k(\xi\cup x)f(x)e_\lambda(e^{u\phi}-1,\xi)\\
&=z^{-1}\int_{\mathbb R^d}f(x)\int_{\Gamma_0}\lambda(d\xi)k(\xi_{-x}\cup0)e_\lambda(e^{u\phi}-1,\xi)\\
&=z^{-1}\int_{\mathbb R^d}f(x)\int_{\Gamma_0}\lambda(d\xi)k(\xi\cup0)e_\lambda(e^{u\phi}-1,\xi),
\end{align*}
from where  equality  (15) 
follows.\quad $\square$

Thus, by Lemma~1, according to Section~2, we expect that the scaled equilibrium  dynamics (with initial distribution $\mu$) converges to the birth-and-death process with  generator
\begin{multline}
(L_0F)(\gamma)=\sum_{x\in\gamma}\frac12\big(C_v\exp[uE(x,\gamma\setminus x)]+C_u\exp[vE(x,\gamma\setminus x)]\big)\\
\times(F(\gamma\setminus x)-F(\gamma)\\
\text{}+\int_{\mathbb R^d}z\,dy\, \frac12\big(C_u\exp[-(1-v)E(y,\gamma)]+C_v\exp[-(1-u)E(y,\gamma)]\big)\\
\times (F(\gamma\cup y)-F(\gamma)) \label{ftftf}
\end{multline}
and the initial distribution $\mu$. In fact, by \cite{KLR},  $(L_0,\mathcal FC_b(C_0(\mathbb R^d),\Gamma)))$ is a Hermitian, non-negative operator in $L^2(\Gamma,\mu)$, and its Friedrichs' extension
$(L_0,D(L_0))$ is  the
generator of a Markov process on $\Gamma$ with {\it c\'adl\'ag\/} paths.

We recall that $L_\varepsilon$ denotes the $L_K$ generator (given by (13)
) scaled by $\varepsilon$. The following theorem states that, at least on an appropriate set of test functions, the operator $L_\varepsilon$ converges to $L_0$ in the $L^2$-norm.

\begin{theorem}\label{yufi}
For each $f\in C_0(\mathbb R^d)$, we have $e^{\langle f,\cdot\rangle}\in D(L_\varepsilon)$ for all $\varepsilon\ge0$, and
$$ L_\varepsilon e^{\langle f,\cdot\rangle}\to L_0
e^{\langle f,\cdot\rangle}\quad\text{in }L^2(\Gamma,\mu)\text{ as }\varepsilon\to0.$$
\end{theorem}

{\it Proof.} We will only sketch the proof of the theorem. Let $f\in C_0(\mathbb R^d)$.
By approximation, one easily shows that, for each $\varepsilon\ge0$, the function
$F(\gamma)=e^{\langle f,\gamma\rangle}$ belongs to
$D(L_\varepsilon)$, and that the action of $L_\varepsilon$ onto $F$ is given, for $\varepsilon>0$ by the  right hand side of (13) 
 in which $a$ is replaced
by $a_\varepsilon$, and for $\varepsilon=0$ by (16)
, respectively.

Denote
\begin{align*}
&(\mathcal L_\varepsilon^-F)(\gamma)=\sum_{x\in\gamma}\int_{\mathbb R^d}dy\, a_\varepsilon(x-y)\\
&\times \exp[uE(x,\gamma\setminus x)-(1-v)E(y,\gamma\setminus x)]
e^{\langle f,\gamma\setminus x\rangle}(1-e^{f(x)}),\\
&(\mathcal L_\varepsilon^+F)(\gamma)=\sum_{x\in\gamma}\int_{\mathbb R^d}dy\, a_\varepsilon(x-y)\\
&\times \exp[uE(x,\gamma\setminus x)-(1-v)E(y,\gamma\setminus x)]
e^{\langle f,\gamma\setminus x\rangle}(e^{f(y)}-1),\\
&(\mathcal L^-_0F)(\gamma)= C_v\sum_{x\in\gamma}\exp[uE(x,\gamma\setminus x)]e^{\langle f,\gamma\setminus x\rangle}(1-e^{f(x)}),\\
&(\mathcal L^+_0F)(\gamma)=  C_u z\int_{\mathbb R^d}dy\, \exp[-(1-v)E(y,\gamma)]e^{\langle f,\gamma\rangle}(e^{f(y)}-1).
\end{align*}
To prove the theorem, it suffices to show that
\begin{align} &\|\mathcal L^\pm_\varepsilon F\|^2_{L^2(\Gamma,\mu)}\to \|\mathcal L^\pm_0 F\|^2_{L^2(\Gamma,\mu)},\notag\\ & (\mathcal L^\pm_\varepsilon F,\mathcal L^\pm_0F)_{L^2(\Gamma,\mu)}\to \|\mathcal L^\pm_0 F\|^2_{L^2(\Gamma,\mu)}
\label{yuyfgdf}\end{align} as $\varepsilon\to0$. To this end, one proceeds as follows. By using (9)
, one represents each of the expressions appearing in  (17) 
 in terms of integrals over $\Gamma$ with respect to $\mu$, as well as integrals over $\mathbb R^d$ with respect to Lebesgue measure. As a result one gets rid of all summations $\sum_{x\in\gamma}$.
Then, one  makes a change of variables, so that instead of $a_\varepsilon(x-y)$ one gets $a(x)$, and in $y$ variable one gets a function which is dominated by an integrable function of $y$. Next, one replaces integration $\int_\Gamma \mu(d\gamma)\dotsm$ by corresponding integration $\int_{\Gamma_0}\lambda(d\eta) k(\eta)\dotsm$. In the obtained expression, one  represents the correlation functional through  a sum of Ursell functionals.
Finally, one takes the limit as $\varepsilon\to0$ by analogy with the final part of the proof of Theorem~\ref{cfttxd}.\quad $\square$

By using the well-known result of the theory of semigroups (see e.g. \cite{Dav}), we get the following corollary of Theorem~\ref{yufi}.

\begin{corollary}\label{yuaxf}
Assume that the set of finite linear combinations of exponential functions $e^{\langle f,\cdot\rangle}$, $f\in C_0(\mathbb R^d)$, is a core for the limiting
generator $(L_0,D(L_0))$. Then, we have the weak convergence of finite-dimensional distributions of the scaled  Markov process in $\Gamma$ with the generator $(L_\varepsilon,D(L_\varepsilon))$ and with the initial distribution $\mu$ to the Markov process in $\Gamma$ with the generator $(L_0,D(L_0))$ and with the initial distribution $\mu$. In particular, if additionally $\phi\ge0$, then this kind of convergence holds when $u=v=0$.
\end{corollary}

We note that the final statement of Corollary~\ref{yuaxf} holds due to a result of \cite{KL} on essential self-adjointness of the generator of Glauber dynamics in the case $\phi\ge0$ and $u=v=0$
(see also \cite{FKL}). In the latter case,  we even expect that the {\it weak convergence of laws\/} holds.  To this end, one needs to consider all processes as taking values in a negative Sobolev space. The tightness of the laws of scaled processes may be
proven by analogy with the proof of \cite[Theorem~7.1]{G}. Next, one shows that this set of laws has, in fact, a unique limiting point---the law of the Markov process with generator $(L_0,D(L_0))$ and initial distribution $\mu_0$.
This is done by identifying the limit via the martingale problem, and using convergence of the generators (compare with the proof of \cite[Theorem~6.7]{GKLR} and that of \cite[Theorem~7.5]{G}).

\begin{center}
{\bf Acknowledgements}
\end{center}

The authors acknowledge the financial support of the SFB 701
`Spectral structures and topological methods in mathematics',
Bielefeld University, and FCT, POCI2010, FEDER.

\end{document}